\documentclass[a4paper]{jpconf}
\usepackage{graphicx}
\usepackage[compress,sort]{cite}
\begin{document}
\title{Chemical and kinetic equilibrations via radiative parton transport}

\author{Bin Zhang and Warner A Wortman}

\address{Department of Chemistry and Physics,
Arkansas State University,
P.O. Box 419, State University,
AR 72467-0419, U.S.A.}

\ead{bzhang@astate.edu}

\begin{abstract}
A hot and dense partonic system can be produced in the early stage 
of a relativistic heavy ion collision. How it equilibrates 
is important for the extraction of Quark-Gluon Plasma properties. 
We study the chemical and kinetic equilibrations of the 
Quark-Gluon Plasma using a radiative transport model. Thermal and 
Color-Glass-Condensate motivated initial conditions are used. 
We observe that screened parton interactions always lead to partial 
pressure isotropization. Different initial pressure anisotropies result 
in the same asymptotic evolution. Comparison of evolutions with and 
without radiative processes shows that chemical equilibration 
interacts with kinetic equilibration and radiative processes can 
contribute significantly to pressure isotropization.
\end{abstract}

\section{Introduction}
The Quark-Gluon Plasma can be produced in the collisions of 
relativistic heavy ions
\cite{Arsene:2004fa,Back:2004je,Adams:2005dq,Adcox:2004mh}. 
Hydrodynamics is very successful in describing relativistic 
heavy ion collisions \cite{Teaney:2000cw,Huovinen:2001cy,Kolb:2001qz,Hirano:2005xf,Heinz:2009xj,Muronga:2001zk,Dusling:2007gi,Luzum:2008cw,Song:2007ux,Song:2008si,Song:2007fn,Denicol:2010xn}. The hydrodynamical 
behavior of the Quark-Gluon Plasma can be investigated with a
microscopic model such as the parton cascade model
\cite{Bass:2002fh,Molnar:2001ux,Zhang:1999bd,Lin:2000cx,Lin:2001yd,Lin:2004en,Xu:2004mz,Ferini:2008he,Xu:2007aa,Zhang:2008zzk,Huovinen:2008te,El:2009vj,Zhang:2009dk,Zhang:2010fx}. We will use a parton
cascade to study the effects of medium dependent cross sections
and radiative processes on the thermalization of a Quark-Gluon
Plasma.

We will focus on a system of gluons. The 2-body collision cross
section, $\sigma_{22}$, is taken to be the perturbative QCD
cross section regulated by a Debye screening mass. The Debye
screening mass squared is proportional to the sum of inverses
of particle momenta. In a dense system, the screening mass is large,
and the cross section is small. This can help avoid many
conceptual and technical problems associated with large
cross sections in dense media. The 2 to 3 cross section, 
$\sigma_{23}$, describes the simplest particle multiplication
process, i.e., two incoming gluons collide and produce three
outgoing gluons. The cross section is taken to be 50\% of the
2 to 2 cross section. This is in line with a more sophisticated
study by Xu and Greiner \cite{Xu:2004mz}. 
All outgoing particles are taken to
be isotropic in the center of mass frame of a collision. 
This is an approximation for the dense system
when screening is important. The inverse process in which there
are 3 incoming gluons and 2 outgoing gluons is completely
determined by the detailed balance relation in order to have
the correct chemical equilibrium. The 3 to 2 collision rate
is determined by the 3 to 2 reaction integral, $I_{32}$, 
i.e., the transition matrix
element modulus squared integrated over the outgoing particle
phase space (with appropriate averaging over initial and 
summing over final internal degrees of freedom). In the
isotropic case, it is directly proportional to $\sigma_{23}$.

\section{Thermalization in a box}

\begin{figure}[h]
\centering
\begin{minipage}{16pc}
\includegraphics[width=16pc]{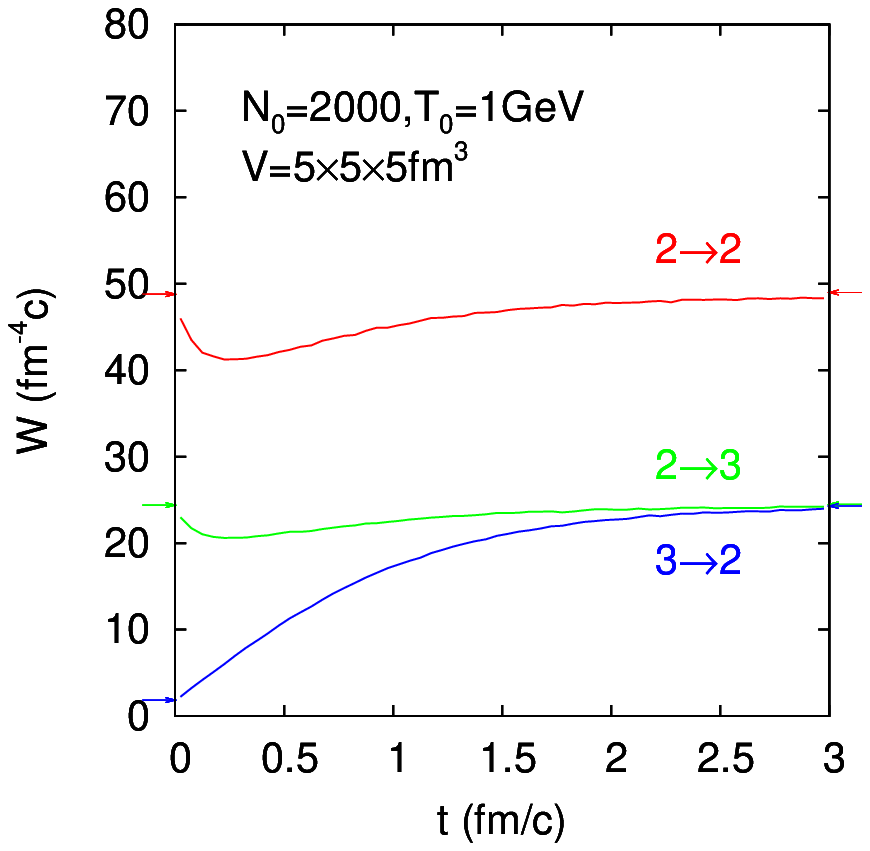}
\caption{\label{fig:box_rate1}Collision rate per unit volume evolutions. 
The arrows indicate initial and equilibrium collision rates per unit volume.}
\end{minipage}\hspace{2pc}%
\begin{minipage}{16pc}
\includegraphics[width=16pc]{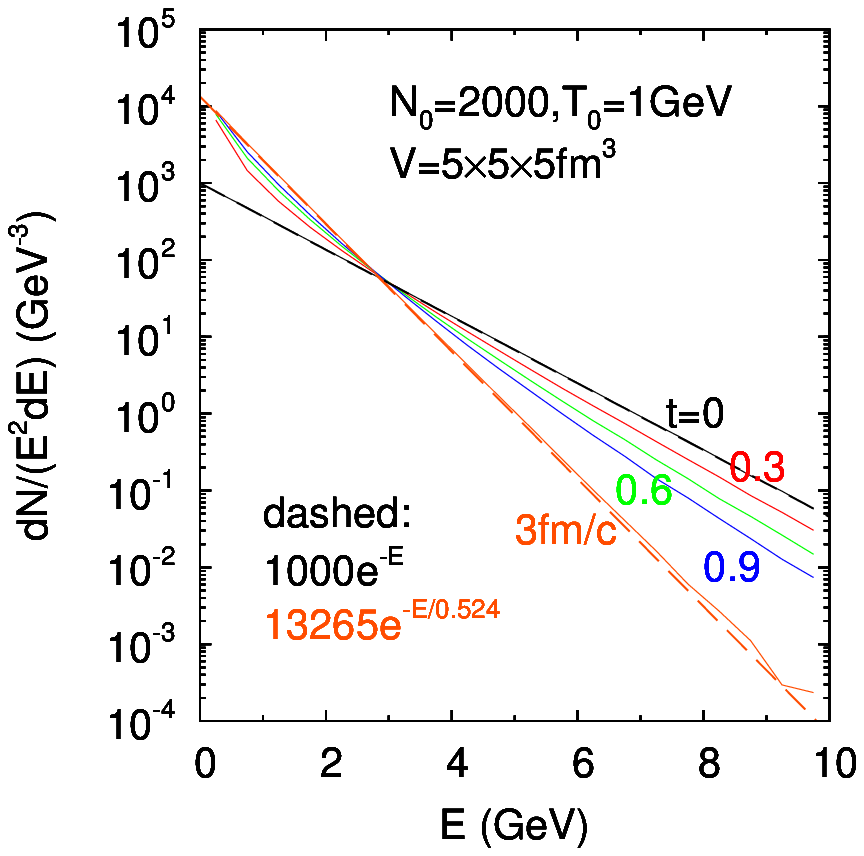}
\caption{\label{fig:box_dndpe1}Particle energy distributions at various times.
The orange dashed line is the distribution in chemical
and kinetic equilibrium.}
\end{minipage} 
\end{figure}

Chemical and kinetic equilibrations of a partonic system can be studied
in a box. For example, we can put 2000 gluons with an initial thermal
distribution characterized by an initial temperature of 1 GeV in a
box of volume $5\times 5\times 5$ fm$^3$. Collision rates per unit
volume as functions of time are plotted in Figure~\ref{fig:box_rate1}.
The $2\rightarrow 2$ rate decreases and then increases toward the
equilibrium value. In this case, the equilibrium rate is identical
to the initial rate and both are determined by the energy density
of the system. As the initial particle number is below the chemical 
equilibrium value, there will be particle production. 
Figure~\ref{fig:box_dndpe1} shows that the production of a large 
number of soft particles makes the energy distributions at early
times out of kinetic equilibrium. In particular, the distribution
at $0.3$ fm/$c$ decreases significantly faster than exponential
in the low energy region.
This leads to an increase in the screening mass and the 
decrease of the cross section. The initial collision rate 
decreases in spite of the particle production and 
the corresponding increase in particle density. The collision rate
gradually relaxes to the equilibrium rate as the momentum distribution
relaxes to the equilibrium exponential distribution. 
Figure~\ref{fig:box_rate1} also shows that the $2\rightarrow 3$ rate 
follows the $2\rightarrow 2$ rate and the $3\rightarrow 2$ rate 
increases steadily and approaches the $2\rightarrow 3$ rate.
At $3$ fm/$c$, the $3\rightarrow 2$ rate is about the same as
the $2\rightarrow 3$ rate and they are both close to the
equilibrium rate. One can conclude that chemical equilibrium is reached
at about $3$ fm/$c$. This is also reflected by the comparison
of the energy distribution at $3$ fm/$c$ with that for the system
in chemical and kinetic equilibrium as shown in 
Figure~\ref{fig:box_dndpe1}.

\section{Thermalization in heavy ion collisions}

\begin{figure}[h]
\centering
\includegraphics[width=16pc]{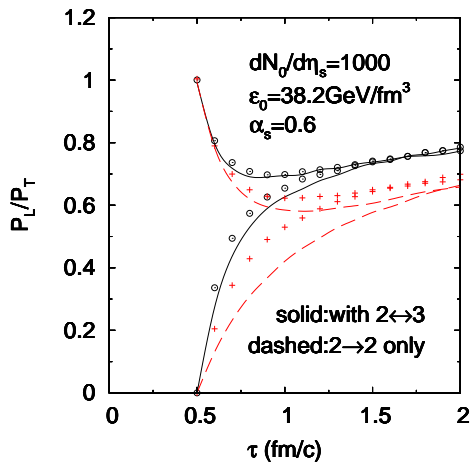} \hspace{2pc}%
\raisebox{6pc}{
\begin{minipage}[b]{16pc}\caption{
\label{fig:plopt1}Pressure anisotropy evolutions. 
The lines are for thermal initial momentum distributions
and the symbols are for Color-Glass Condensate motivated 
initial momentum distributions. The solid lines and circles
include the $2\leftrightarrow 3$ processes while the dashed
lines and pluses have only elastic collisions.}
\end{minipage}
}
\end{figure}

Thermalization in heavy ion collisions has its unique aspects
as compared to thermalization in a box. We will focus on
the central cell during the early stage in central collisions
when the system can be considered to undergo only longitudinal
expansion. Kinetic equilibration can be characterized by 
the longitudinal to transverse pressure ratio, $P_L/P_T$,
which is also called the pressure anisotropy. When the 
system is in kinetic equilibrium, the pressure anisotropy
equals $1$. A pressure anisotropy different from $1$ signals
that the system is not in equilibrium. Different from 
thermalization in a box, longitudinal expansion
tends to decrease the pressure anisotropy. Even when
the initial pressure anisotropy equals $1$, free 
streaming can lead to a decrease of the pressure anisotropy
toward $0$. Particle interactions can counteract the
effects of expansion. In particular, a medium dependent
cross section increases with time when the system expands
and the density decreases. This can lead to pressure
isotropization as shown in Figure~\ref{fig:plopt1}. For
example, the black solid line starts with an initial pressure
anisotropy of $1$ and expansion dominates over equilibration
initially and the pressure anisotropy decreases with time.
As time goes on, equilibration wins over expansion and the
pressure anisotropy stops decreasing and eventually turns
toward $1$.

How important are the radiative
processes? Comparison of the solid lines and dashed lines
in Figure~\ref{fig:plopt1} shows that radiative processes
can significantly enhance thermalization. The 
initial energy density and particle rapidity are about
the same as those expected at RHIC and around chemical
equilibrium. Therefore, radiative processes are important
for thermalization in relativistic heavy ion collisions.
If the system starts with a different pressure anisotropy,
e.g., $P_L/P_T=0$, as shown by the lower solid curve in
Figure~\ref{fig:plopt1}, after some time, the memory of
the initial anisotropy is lost and the evolution converges
with that from $P_L/P_T=1$ toward a common asymptotic evolution. 
In addition to the initial pressure anisotropy,
the initial momentum distribution may affect the 
thermalization of the system. This can be studied by 
comparing results starting from exponential
initial momentum distributions and those from 
step function initial momentum distributions. 
An exponential distribution mimics a thermal distribution,
while a step function is a simplification of the 
Color-Glass-Condensate initial condition where
the distribution is approximately Bose-Einstein
below the saturation scale and has a power-law behavior
above the saturation scale \cite{Mueller:1999fp,
Mueller:1999pi,Bjoraker:2000cf,El:2007vg}. It is interesting
to see from Figure~\ref{fig:plopt1} that for the
elastic only case, the step function initial momentum
distribution significantly increases isotropization while
there is not much change for the radiative case. This
is caused by the slow thermalization of the elastic
only case. It maintains the step function shape for
a longer period of time compared to the radiative case
and hence has smaller screening mass and larger cross
section compared to the evolution from an exponential
initial momentum distribution.

The pressure anisotropy evolution depends on system
parameters such as the coupling constant $\alpha_s$ 
and the initial energy density $\epsilon_0$. 
When radiative processes
are included, the evolution depends sensitively on the
coupling constant and is not very sensitive to the
initial energy density. However, the elastic only case
depends on both the coupling constant and initial
energy density. There is an approximate $\alpha_s$ scaling
for the radiative case, qualitatively different from
the exact $\alpha_s\epsilon_0$ scaling observed in
the elastic only case. More discussions on this
can be found in \cite{Zhang:2010fx}.

\begin{figure}[h]
\centering
\begin{minipage}{16pc}
\includegraphics[width=16pc]{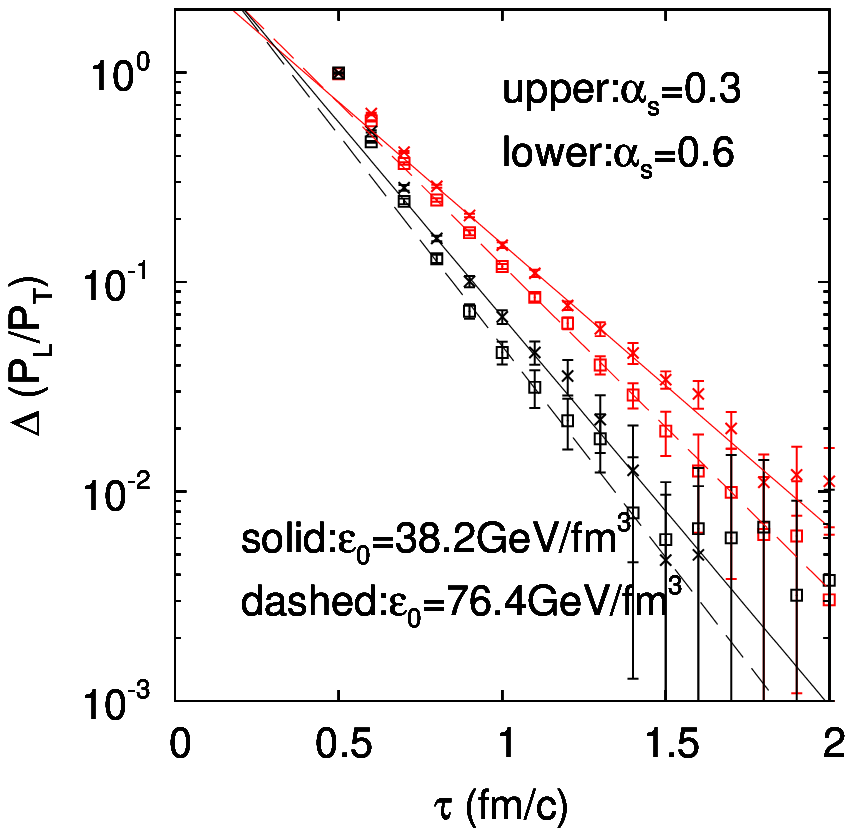}
\caption{\label{fig:dplopt1}Pressure anisotropy difference evolutions
with radiative processes.}
\end{minipage}\hspace{2pc}%
\begin{minipage}{16pc}
\includegraphics[width=16pc]{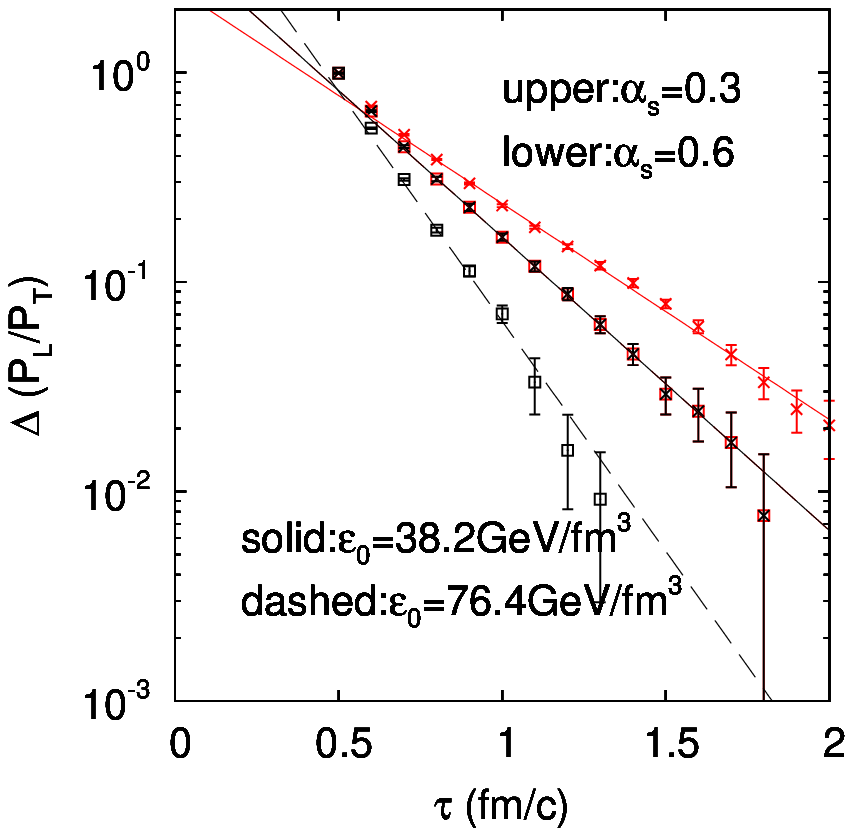}
\caption{\label{fig:dplopt2}Pressure anisotropy difference evolutions
with only elastic collisions.}
\end{minipage} 
\end{figure}

The pressure anisotropy evolution loses memory of the initial
pressure anisotropy. This can be studied more carefully by
looking at the difference between the evolutions for systems
with different initial pressure anisotropies but otherwise
identical initial setups. The pressure anisotropy 
difference evolutions are plotted in Figure~\ref{fig:dplopt1} 
and Figure~\ref{fig:dplopt2} for the radiative and elastic only
cases respectively. These two figures show that after a
very short period of time, the pressure anisotropy evolution
becomes exponential. The larger the coupling constant or
the larger the initial energy density, the faster the decrease
toward zero is. Figure~\ref{fig:dplopt2} demonstrates
the $\alpha_s\epsilon_0$ scaling again as the upper dashed
curve overlaps the lower solid curve. It is also interesting
to note that when the evolutions are traced back in time, 
they appear to have come from the same point, even though 
the point for the radiative case appears to be quite different 
from the point for the elastic only case. 

\begin{figure}[h]
\centering
\includegraphics[width=16pc]{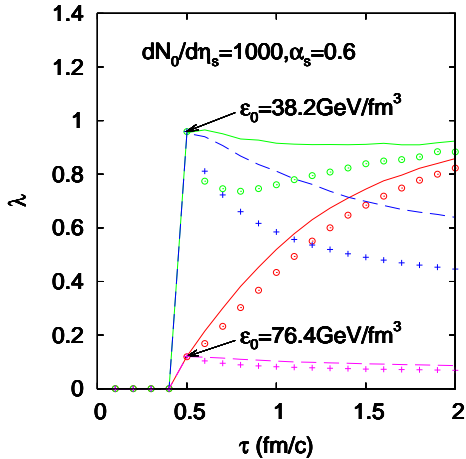} \hspace{2pc}%
\raisebox{7pc}{
\begin{minipage}[b]{16pc}\caption{
\label{fig:lambda1}Fugacity evolutions. 
Red and Green are for the radiative case while blue and magenta
are for the elastic only case. Evolutions from isotropic initial
conditions are shown as lines and those from transverse initial
conditions are shown as symbols.}
\end{minipage}
}
\end{figure}

Chemical equilibration can be characterized by the fugacity evolution. 
Figure~\ref{fig:lambda1} gives information on how chemical
equilibration interacts with kinetic equilibration. For the
radiative case, even systems starting from
the same initial fugacity can have
different fugacity evolutions depending on the initial pressure
anisotropy. Isotropic initial conditions are more effective in
maintaining chemical equilibrium as seen from the green solid
line while transverse initial conditions have slower chemical
equilibrations as the green circles show. On the other hand,
The two solid lines have very different chemical equilibrations
but almost the same pressure anisotropy evolution. The 
interplay between kinetic equilibration and chemical 
equilibration shows that they are two inseparable aspects
of the thermalization process.

\section{Summary and outlook}
When there is chemical equilibration (radiative processes), 
kinetic equilibration (pressure anisotropy evolution) depends 
mainly on the coupling constant. The memory of the initial 
pressure anisotropy is quickly lost and the asymptotic 
evolution is not sensitive to the initial pressure anisotropy. 
The pressure anisotropy evolution is not sensitive to changes 
in the initial momentum distribution or the initial 
energy density (or fugacity) either. The schematic 
model used here can certainly be made more realistic in 
many ways and be applied to the study of the effects of 
radiative processes on many experimental observables.

\ack
We would like to thank J. Aichlin, S. Bass, R. Bellwied, 
F. Gelis, H. Grigoryan, M. Guenther, U. Heinz, S. Katz, 
V. Koch, M. Lisa, L. McLerran, H. Meyer, U. Mosel, 
J.-Y. Ollitrault, S. Pratt, I. Vitev, H. Zhou
for helpful discussions and
the Parallel Distributed Systems Facilities of the
National Energy Research Scientific Computing Center
for providing computing resources.
This work was supported by the U.S. National Science Foundation
under grant No.'s PHY-0554930 and PHY-0970104.

\end{document}